\documentclass[twocolumn,showpacs,showkeys,prb,footinbib,aps]{revtex4-1}
\usepackage{hyperref}  
\usepackage{graphicx}
\usepackage{dcolumn}
\usepackage{amsmath}
\begin{document}

\title{A simple and versatile analytical approach for planar metamaterials}
\date{\today}
\author{J\"org Petschulat}
\email{joerg.petschulat@uni-jena.de}
\author{Arkadi Chipouline}
\author{Andreas T{\"u}nnermann}
\altaffiliation[also with: ]{Fraunhofer Institute of Applied Optics and Precision Engineering Jena, Germany.} 
\author{Thomas Pertsch}
\affiliation{Institute of Applied Physics, Friedrich-Schiller-Universit{\"a}t
Jena, Max Wien Platz 1, 07743 Jena, Germany}
\author{Christoph Menzel}
\author{Carsten Rockstuhl}
\author{Thomas Paul}
\author{Falk Lederer}
\affiliation{Institute of Condensed Matter Theory and Solid State Optics,
Friedrich-Schiller-Universit{\"a}t Jena, Max Wien Platz 1, 07743 Jena,
Germany}
\begin{abstract}

We present an analytical model which permits the calculation
of effective material parameters for planar metamaterials consisting of arbitrary 
unit cells (metaatoms) formed by
a set of straight wire sections of potentially different shape.
The model takes advantage of resonant electric dipole oscillations in the wires and their mutual
coupling. The pertinent form of the metaatom determines the actual coupling features.
This procedure represents a kind of building block model for quite different metaatoms.
Based on the parameters describing the
individual dipole oscillations and their mutual coupling the entire effective
metamaterial tensor can be determined. By knowing these parameters for
a certain metaatom it can be systematically modified to
create the desired features. Performing such modifications effective material properties as well as the far field intensities remain predictable. As an example the model is applied
to reveal the occurrence of optical activity if the split ring resonator
metaatom is modified to L- or S-shaped metaatoms.
\end{abstract}


\maketitle

\section{Introduction}
Metamaterials extend the optical effective material response of natural
media. They are artificial materials that allow to tailor light propagation
properties by a careful design of the mesoscopic metaatoms the
metamaterial is made of. By controlling the geometry and selecting the
material dispersion of the metaatom, novel effects such as negative
refraction \cite{Shalaev2007,Wegener2007,Zhang2008}, optical cloaking
\cite{Zhang2009,Engheta2009,Lai2009,Farhat2009,Pendry2006,Leonhardt2006},
as well as a series of optical analogues to phenomena known from
different disciplines in physics have been observed
\cite{Narimanov2009,Kivshar2009,Guney2009,Zheludev2008,Giessen2009}.

In principle, the material response is mathematically best described in
terms of constitutive relations; leading to tensors that relate the electric
displacement and the magnetic induction to the electric and magnetic
fields, respectively. In the initial stage of research on metamaterials
emphasize was put on exploring materials that potentially lead to a
bi-axial anisotropic (linear dichroism) effective material response
\cite{Shalaev2007,Wegener2007,Helgert2009,Zhang2008,Garcia2009,Wegener2009}.
Recently research was also extended towards the exploration of
metaatoms that affect off-diagonal elements of the effective material
tensors (elliptical dichroism). It expands the number of observable optical
phenomena, leading to, e.g., optical
activity\cite{Bai2007,Arnaut1997,Reyes2006,Zheludev2009_2,Tretyakov2003},
bidirectional and asymmetric transmission
\cite{Zheludev2006,Zheludev2007,Zhukovsky2009} or chirality-induced
negative refraction \cite{Pendry2004,Tretyakov2005,Soukoulis2009}.

In our understanding optical activity comprises all effects on light
propagation resulting from the nonlinearity of the polarization
eigenstates, hence including phenomena like circular dichroism, all
effects of genuine three-dimensional chirality and the optical
manifestations of planar chirality \cite{Zhukovsky2009}.

However, despite the option to resort to rigorous computations for
describing the light propagation on the mesoscopic level of the
metaatoms, an enduring issue in metamaterial research is the question
how the effective material tensor looks like for a certain metamaterial.

In general, investigating the geometry of the metamaterial (the
metaatoms geometry \textit{and} their arrangement) allows to determine
the form of the effective material tensors in the quasi-static limit as
extensively discussed in Ref. \onlinecite{Arnaut1997}. From such
considerations it is possible to conclude on the symmetry of the
plasmonic eigenmodes sustained by the metaatoms and on the
polarization of the eigenmodes allowed to propagate in the effective
medium \cite{Zheludev2006}. But in order to determine the actual
frequency dependence of the tensor elements, more extended models are
needed which start in their description of the metaatom properties from
scratch \cite{Zhukovsky2009}. Such models are required to be universal,
simple and assumption-free to the largest possible extent.

Here we outline an approach which meets these requirements. It is based on
conceptually decomposing  the complex metatoms into a set of coupled
plasmonic entities that sustain the excitation of dipolar resonances
\cite{Petschulat2008}. The knowledge of the plasmonic properties of these
dipoles and their coupling suffices to derive the material response and the
symmetry of the eigenmodes. This, in turn, permits to predict the observable
quantities in the far-field, such as the polarization and frequency-dependent
reflected and transmitted complex amplitudes.

The most appealing aspect of our model is that once the plasmonic entities and
their coupling strengths are characterized, far-field properties remain
predictable by the analytical model even if substantial modifications of the
metaatom geometry have been made. Even a modification that leads to a
different symmetry of the material tensor does not prevent a quantitative
description of the dispersive behavior of the tensor elements. To become
specific, we start our investigations with an optical inactive, bi-axial anisotropic
metamaterial, namely the split-ring resonator (SRR). From the observable
far-field quantities we derive the properties of the dipolar oscillators that govern
the plasmonic response of the three wires forming the SRR, namely the
individual eigenfrequencies, oscillator strengths and damping constants as well
as their mutual coupling strengths. By relying on these quantities, we determine
the effective properties of metamaterials consisting of modified metaatoms with
respect to the initial SRR.

We focus on two modifications that evoke optical activity. With our analytical
model we predict the effective properties of these metamaterials. We then
compute the optical coefficients for a slab made of these metamaterials and
compare it to rigorous simulations. Excellent agreement between the analytically
predicted and the rigorously calculated optical coefficients is observed
throughout our work. Therefore, the proposed method can be used for the
parameter retrieval without resorting to rigorous simulations. And since it is
based on analytical considerations the approach potentially allows a large variety
of metaatom modifications and to systematically tailor its effective properties.
Hence our approach provides a powerful and versatile tool for a systematic
analysis of achievable material properties by varying only a few constituents that
may couple in some well-defined ways.

Furthermore, we will also show that such an analytical treatment provides
further insight into metamaterial properties. Specifically, we show that it is
possible to directly infer that the model's predictions are valid in terms of the
Casimir-Onsager relations \cite{Onsager1931,Casimir1945}, the requirement for
time-reversal and reciprocity in linear media \cite{Tretyakov2002}.

Thus, the main aspects of the paper can be summarized as follows. At first, the
localized carrier dynamics occurring in metaatoms may be properly described by
a set of coupled oscillators, representing the decomposition of the metaatom in
nanowire pieces. Second, the dynamics of these oscillators, determined by the
shape of the nanowires and their coupling, result in electric dipole polarizabilities
that permit the calculation of the effective permittivity tensor. The main
advantage of this simple approach is that modifications of a metaatom, for
which the oscillator dynamics and parameters have been found, leave the
effective permittivity tensors predictable. Hence, the far field reflectance and
transmittance can be calculated. We will show that this holds also for
modifications changing the character of the eigenstates from linear to non-linear
polarized. Moreover, the approach can be useful to determine the effective
permittivity tensor for optically active metamaterials, whose eigenstates are in
general elliptically polarized, by intensity measurements of linearly polarized
light.

\section{The planar SRR metaatom}

\begin{figure}
\includegraphics[width=8.7cm]{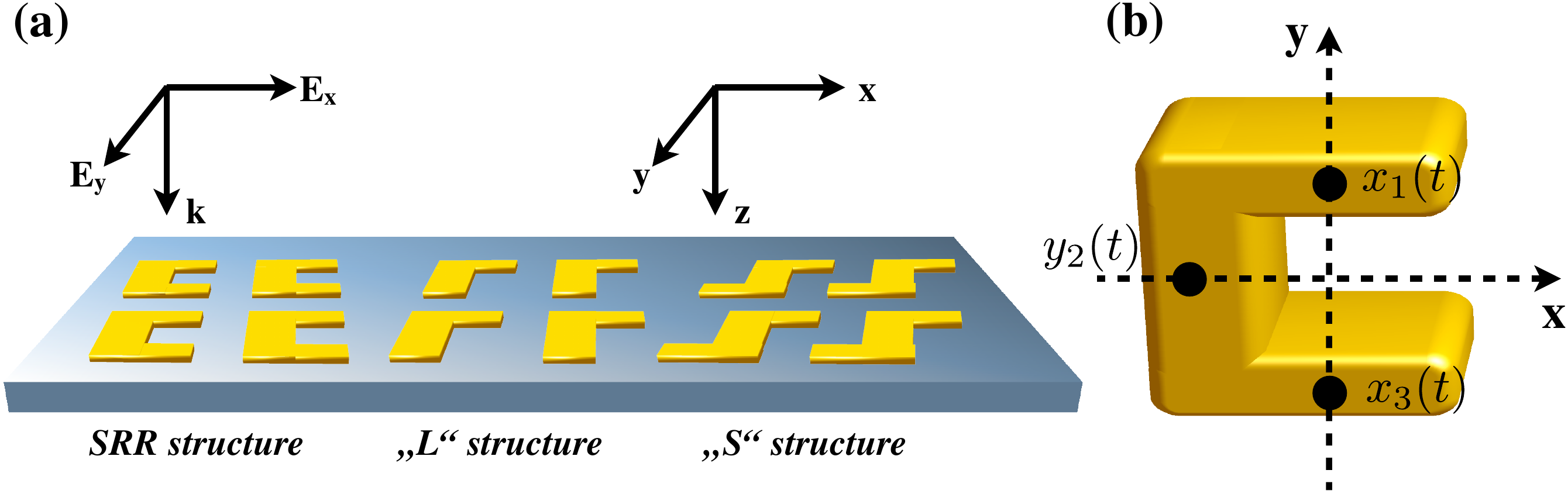}
\caption{(Color online) (a) The original SRR structure (left), the first
modification, namely the L- (center) and the second
modification, the S-structure (right). (b) The SRR
together with the carrier oscillators, marked by black
dots, which are used to phenomenologically replace the SRR.} \label{pic1}
\end{figure}

To reveal the versatile character of our approach we will directly start by
conceptually replacing the planar SRR geometry, shown in Fig. \ref{pic1}(b),  by a
set of coupled oscillators. Each of the oscillators introduced here represents a
metaatoms piece, i.e. a straight nanowire that is coupled to its electrically
conductive neighbors. The oscillators accounting for the isolated nanowire are
associated with the excitation of carriers representing the free-electron gas of
the metal. Hereby the carrier dynamics is solely influenced by the external
electromagnetic field, including contributions from adjacent nanowires. The
dynamics becomes resonant for the eigenfrequency of the localized plasmon
polariton resonance of the individual nanowire. This localized resonance
described by the individual oscillator corresponds to the fundamental electric
dipole mode since the dimensions of the nanowires forming the metaatom
investigated here are small compared with the wavelength. We will show later
that the carrier oscillations of metaatoms assembled from several coupled wires
still correspond to electric dipole polarizabilities. This holds as long as the wires
are assembled in-plane. For out-of-plane structures higher order multipoles
come generally into play
 \cite{Petschulat2008,Petschulat2009}.

Here the spatial coordinate represents the elongation of a negatively
charged carrier density driven by an external electromagnetic field, which
is the usual assumption in plasmonics \cite{RaetherBook}.

We show below that this assumption is sufficient to entirely predict the
optical response. The equations for the coupled oscillators are
\begin{eqnarray}
\label{Eq1}
\frac{\partial^2}{\partial t^2}x_1+\gamma_1\frac{\partial}{\partial t}x_1+\omega_{01}^2 x_1+\sigma_{21}y_2=-\frac{q_1}{m}E_x  , \nonumber \\
\frac{\partial^2}{\partial t^2}y_2+\gamma_2\frac{\partial}{\partial t}y_2+\omega_{02}^2 y_2+\sigma_{21}x_1-\sigma_{23}x_3=-\frac{q_2}{m}E_y , \nonumber \\
\frac{\partial^2}{\partial t^2}x_3+\gamma_3\frac{\partial}{\partial t}x_3+\omega_{03}^2 x_3-\sigma_{23}y_2=-\frac{q_3}{m}E_x, \nonumber\\
\end{eqnarray}
where we have assumed nearest neighbor coupling between adjacent,
i.e. conductively coupled dipolar oscillators. The oscillators are driven by
the electric field component of an external illuminating field propagating
in $z$ direction (normal incidence is assumed throughout the entire
manuscript). Excitation of the oscillators by magnetic field components
can be safely neglected \cite{TretyakovBook,Petschulat2008}. In Eqs.
(\ref{Eq1}) $\gamma_j$

is the damping, $\omega_{0j}$ the eigenfrequency, $\sigma_{ij}$ the coupling
constant and $q_j$ is the charge for the three oscillators $(i,j) \in (1,2,3)$,
respectively,  similarly to those introduced in Ref. \onlinecite{Petschulat2008}.
The coordinates $(x_1,y_2,x_3)$ themselves are understood as the
displacement of the negatively charged carriers representing oscillating currents
or multipole moments where both approaches can be used to determine the
effective material response of an artificial medium exhibiting a carrier dynamics
as described by Eq. (\ref{Eq1}) \cite{Pershan1963}. Here we apply the multipole
approach to map the displacement onto electric dipole moments, since this is
consistent with our previous works \cite{Petschulat2008,Petschulat2009}.

Accounting for the contributions of the electric dipole and quadrupole as well as
the magnetic dipole moment the wave equation reads as
\begin{eqnarray}
\label{Eq3}
\Delta\textbf{E}(\textbf{r},\omega)=-\mu_0 \epsilon_0\omega^2\textbf{E}(\textbf{r},\omega)
-\omega^2\mu_0\textbf{P}(\textbf{r},\omega) \nonumber \\
+\omega^2\mu_0\nabla\cdot\textbf{Q}(\textbf{r},\omega)
-i\omega\mu_0\nabla\times\textbf{M}(\textbf{r},\omega),
\end{eqnarray}

We put emphasis on the fact that the higher order multipole moments,
leading to $\mathbf{Q},\mathbf{M}$,  appear in general \cite{Giessen2009b}, but
do not provide any contribution in the present configuration, as mentioned above. Thus, there
is no effective magnetic response (effective permeability tensor) as
expected and it is sufficient to consider the dispersion in the
susceptibility, resulting in a linear effective permittivity tensor only
\cite{Rockstuhl2007}. This will hold for all planar configurations with an
illuminating field  invariant in the $x-y$-plane, hence no spatial
dispersion occurs which would result in an artificial magnetic response.
Substituting the displacements (\ref{Eq1}) into the definition of the dielectric polarization
\cite{RaabBook} we can introduce the effective susceptibility tensor
$\hat{\mathbf{\chi}}(\omega)$
\begin{eqnarray}
\label{Eq2}
\textbf{P}(\textbf{r},t)&=&\eta\sum_{l=1}^Nq_l\textbf{r}_l,  \nonumber \\
P_{i}(\textbf{r},\omega)&=&\epsilon_0\chi_{ij}(\omega)E_{j}(\textbf{r},\omega),
\end{eqnarray}
where  $\eta$ accounts for the carrier density.
This effective susceptibility can be easily calculated.  As usual we define the effective permittivity tensor as
\begin{equation}
\hat{\mathbf{\epsilon}}(\omega)=\underline{\mathbf{1}}+\hat{\mathbf{\chi}}(\omega),
\label{Eq4}
\end{equation}
that governs
the wave propagation of an
incident plane wave in an effective medium composed of SRR
metaatoms.

Next we consider the possible eigenmodes of Eqs. (\ref{Eq1}) for the two
polarization directions ($x$ or $y$). It suffices to investigate
these two polarizations as long as the system under consideration is
linear. In order to describe a SRR with two identical side arms we set
$\omega_{01}=\omega_{03}\equiv\omega_{0x},~\gamma_1=\gamma_3\equiv{\gamma_x},~\sigma_{21}=\sigma_{23}\equiv\sigma$
and $q_1=q_3\equiv -q_x$. For the oscillator associated with the SRR
base having a different geometry, we put
$\omega_{02}\equiv\omega_{0y},~\gamma_2\equiv\gamma_y$ and
$q_2\equiv -q_y$. We note that the latter distinction could have been
dropped if the geometry of all constituents of the SRR would have been
the same. For the reasons obvious from the consideration below we
refrain from doing so.

For a polarization of the incoming plane wave parallel to the $x$-axis we
can solve Eqs. (\ref{Eq1}) and obtain the following displacements
in Fourier domain:
\begin{eqnarray}
\label{Eq5}
x_1(\omega)&=&x_3(\omega)=-\frac{q_x}{m}\frac{1}{A_{x}(\omega)}E_x(z,\omega), \nonumber \\
y_2(\omega)&=&0,
\end{eqnarray}
where we have introduced
$A_{x}(\omega)=\omega_{0x}^2-\omega^2-i\omega\gamma_x$.

For the polarization in $y$-direction we obtain
\begin{eqnarray}
\label{Eq6}
x_1(\omega)&=&-x_3(\omega)=-\frac{q_y}{m}\frac{\sigma}{A_{x}(\omega)A_y(\omega)-2\sigma^2}E_y(z,\omega), \nonumber \\
y_2(\omega)&=&-\frac{q_y}{m}\frac{A_x(\omega)}{A_{x}(\omega)A_y(\omega)-2\sigma^2}E_y(z,\omega),
\end{eqnarray}
with $A_{y}(\omega)=\omega_{0y}^2-\omega^2-i\omega\gamma_y$.

\begin{figure}
\includegraphics[width=8.7cm]{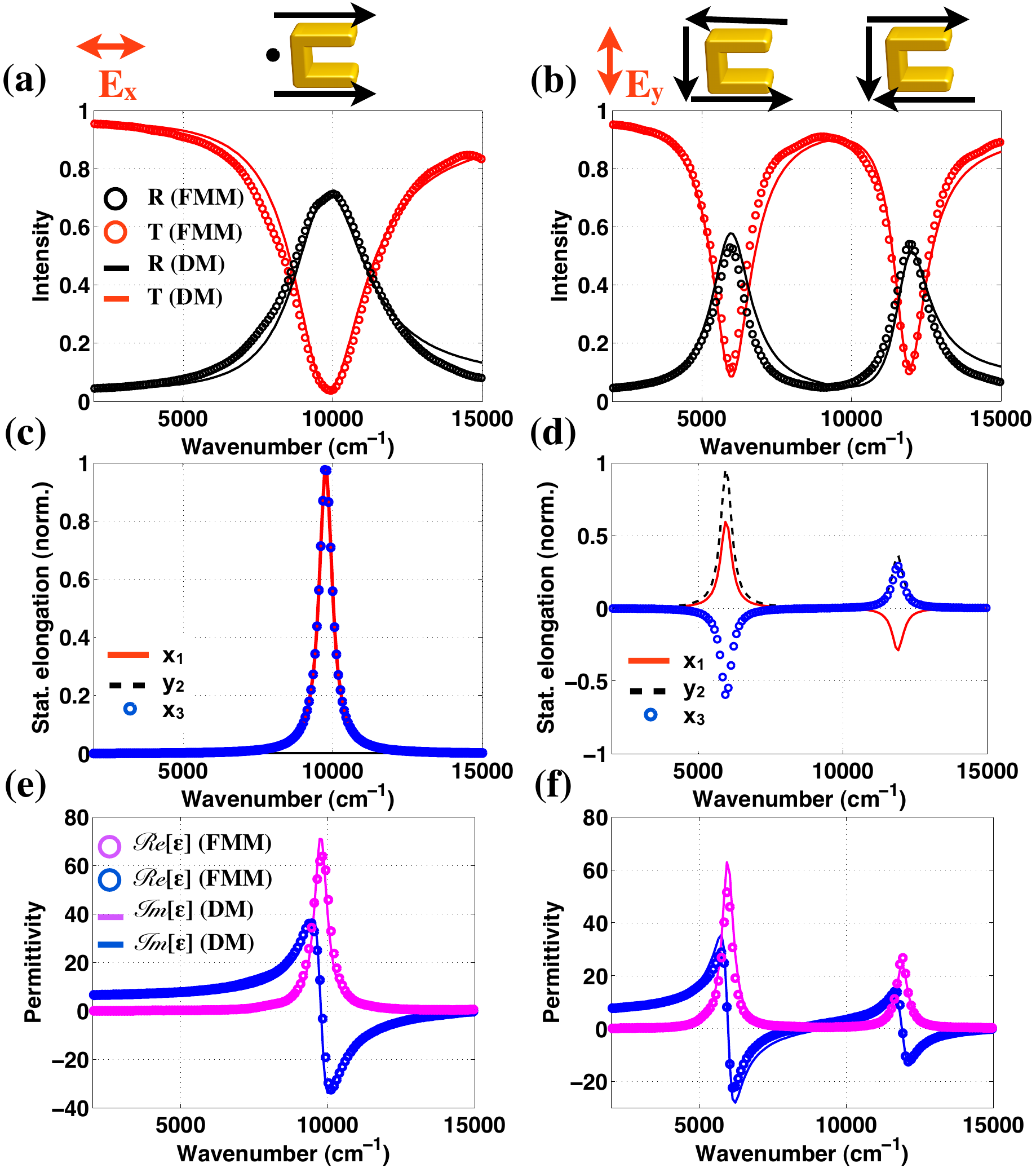}
\caption{(Color online) The rigorously calculated (FMM) far-field transmission/reflection spectra compared with those obtained by the analytical coupled dipole model (DM) for the two indicated polarization directions [(a) $x$-polarization and (b) $y$-polarization]. The stationary carrier elongation (normalized imaginary part of $x_{1,3}$ and $y_2$) for the two corresponding polarizations [(c) and (d)]. (e),(f) The exactly retrieved parameters compared to the analytical calculations.}
\label{pic2}
\end{figure}

Considering the eigenmodes for $x$-polarization, we observe that two
parallel dipoles ($x_1=x_3$) are induced, while due to symmetry
constraints no dipole is induced in $y$-direction ($y_2=0$), see Fig. \ref{pic2}(c).
By contrast, besides a dipole in $y$-direction the $y$-polarized
illumination induces oscillating dipoles in $x$-direction in both arms. But
due to the anti-symmetric oscillation $x_3=-x_1$ [Eq. (\ref{Eq6})], the
dipoles in the SRR arms [Fig. \ref{pic2}(d)] do not radiate into the far-field
because they oscillate $\pi$ out-of-phase and interfere destructively.
Hence, no cross-polarization is observed and the far-field polarization
equals that of the illumination. Later we will prove that any radiation
emerging from cross-polarized dipole moments will result in an optical
activity, as expected.  

By substituting Eqs. (\ref{Eq5}, \ref{Eq6}) in Eqs. (\ref{Eq2}) we get the
susceptibility tensor for the pertinent SRR configuration
\begin{eqnarray}
\label{Eq7}
\hat{\chi}(\omega)&=&\left(\begin{array}{ccc} \chi_{xx}(\omega) & 0 & 0\\ 0 & \chi_{yy}(\omega) & 0\\ 0 & 0 & 0\\ \end{array}\right), \nonumber \\
\chi_{xx}(\omega)&=&\frac{q^2_x\eta}{m\epsilon_0}\frac{2}{A_x(\omega)}, \nonumber \\
\chi_{yy}(\omega)&=&\frac{q^2_y\eta}{m\epsilon_0}\frac{A_x(\omega)}{A_x(\omega)A_y(\omega)-2\sigma^2},
\end{eqnarray}
with the polarization $\textbf{P}=-\eta~(q_x(x_1+x_3),q_yy_2,0)^T$, according to Eq. (\ref{Eq2}). As expected, the
susceptibility tensor is diagonal. Hence, the eigenmodes of the
effectively homogenous medium are linearly polarized and orthogonal.
Due to the polarization dependent carrier dynamics the SRR shows a
linear dichroitic behavior. Hence, our model correctly predicts the linear
polarization eigenstates as required by the mirror-symmetry with respect
to the $x-z$-plane. The only unknown parameters are those describing
the oscillators and their coupling strengths. They can be determined by
matching the optical coefficients of an effective medium whose
permittivity is described by Eqs. (\ref{Eq7}) to the spectra obtained by
rigorous simulations or far-field measurements of the structure.

In order to find the oscillator parameters, we performed numerical Fourier
Model Method (FMM) calculations \cite{Lie1997} of a periodic array of
gold SRR's \footnote{The period in $x$- and $y$-direction is $0.4~\mu
m$, the SRR arm length $0.2~\mu m$, the base width $0.08~\mu m$, the
arm width $0.04~\mu m$ and the metal film thickness $0.025~\mu m$.
Gold material parameters were taken from literature \cite{JC1972}. As a
substrate index we used $n_\text{sub}=1.5$ and for the ambient material
$n_\text{amb}=1$.} similar to those reported in Ref. \onlinecite{Wegener2007}.
These numerical far-field observables (reflected and transmitted
intensities) were fitted by using the effective permittivity tensor Eq.
(\ref{Eq4}) for both polarization directions in a conventional transfer
matrix formalism that computes reflection and transmission from a slab
of equal thickness \cite{Yeh1998}. Results are shown in Fig. \ref{pic2}(a)
and \ref{pic2}(b).

Once the unknown parameters have been found the frequency dependent
stationary elongations of the oscillators can be determined as shown in
Figs. \ref{pic2}(c) and (d). They can be used to identify the carrier
oscillations of the different plasmonic eigenmodes. These carrier
oscillations are shown on top of Fig. (\ref{pic2}) and their horizontal position relates
to the respective resonance frequency. The observed eigenmodes are
documented in literature \cite{Rockstuhl2007}.
Figures \ref{pic2}(e) and (f) show the rigorously retrieved effective material properties
 together with the ones of the analytical model.
Excellent agreement is observed. The rigorous results were obtained by
applying a common parameter retrieval based on an inversion of the
matrix formalism to calculate the effective parameters on the basis of
complex reflected and transmitted amplitudes for a certain slab
thickness. We underline that the unknown oscillator parameters can be
obtained by comparing the far-field intensity only. The derived effective
material parameters are in excellent agreement with the rigorous results,
but they were derived without the necessity of knowing the complex-valued fields. 
This will be advantageous for optical active structures
where the experimental determination of these parameters is in general
involved (e.g. phase resolved measurements are not required).

\section{The L - metaatom}
As outlined in the introduction low-symmetry metaatoms are investigated
in order to reveal new optical phenomena like asymmetric transmission.
We will therefore extend the previous considerations towards optically
active effective media by rearranging the SRR constituents. In the
following we shall rely on the oscillator parameters obtained by the
fitting procedure above. Using these parameters in conjunction with the
analytical expressions for the rearranged constituents as derived below,
the optical response of the modified metaatoms can be predicted.
\begin{figure}
\includegraphics[width=8.7cm]{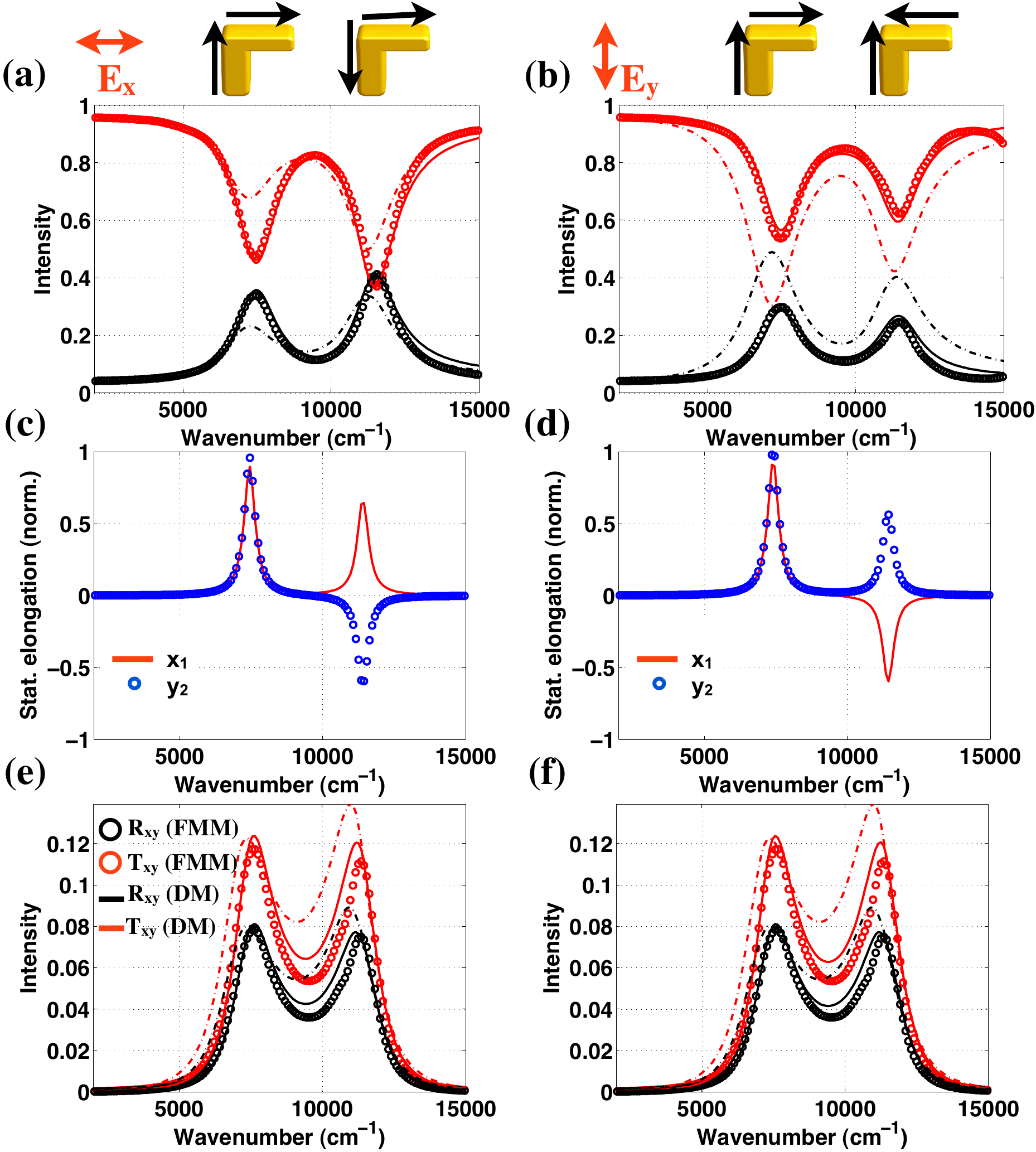}
\caption{(Color online) The far-field response of the L - structure  for  $x~-$ (a) and
$y~-$ polarization (b) . In addition to the numerical (FMM, spheres) and the fitted data (DM, solid lines) the predicted spectra incorporating the SRR parameters (dashed-dotted lines) are plotted.
(c),(d) In contrast to the SRR both
eigenmodes can be excited for each polarization direction. The
respective numerical cross-polarization contributions (FMM, circles) compared with
the model predicted (dashed-dotted lines) and the fitted (solid lines) values are shown in (e) and (f). Note that figures (e) and (f) are identical as required for such kind of effective
media and are only shown for completeness.} \label{pic3}
\end{figure}
The first investigated structure is the L -  metaatom
\cite{Canfield2005,Canfield2005_2}. In this geometry one of the SRR
arms is omitted in order to prevent the cancelation of the far-field that
originated from the antiparallel dipole moments. Hence, it is expected
to obtain polarization rotation.  To get specific in Eq. (\ref{Eq1}) one horizontal
oscillating dipole, e.g., dipole '3', has to be dropped. Hence, we obtain for $x$-polarization
\begin{eqnarray}
\label{Eq8}
x_1^x&=&-\frac{q_x}{m}\frac{A_y(\omega)}{A_x(\omega)A_y(\omega)-\sigma^2}E_x(z,\omega) , \nonumber \\
y_2^x&=&-\frac{q_x}{m}\frac{\sigma}{A_x(\omega)A_y(\omega)-\sigma^2}E_x(z,\omega),
\end{eqnarray}
and for $y$-polarization
\begin{eqnarray}
\label{Eq9}
x_1^y&=&-\frac{q_y}{m}\frac{\sigma}{A_x(\omega)A_y(\omega)-\sigma^2}E_y(z,\omega) , \nonumber \\
y_2^y&=&-\frac{q_y}{m}\frac{A_x(\omega)}{A_x(\omega)A_y(\omega)-\sigma^2}E_y(z,\omega).
\end{eqnarray}
From the resulting polarization Eq. (\ref{Eq2})
\begin{eqnarray}
\label{Eq10}
\textbf{P}^j(z,\omega)=-\eta\left(\begin{array}{ccc}q_xx_1^j\\q_yy_2^j\\0 \end{array}\right)~,~j\in[x,y]
\end{eqnarray}
the susceptibility reads as
\begin{eqnarray}
\label{Eq11}
\hat{\chi}(\omega)&=&\left(\begin{array}{ccc} \chi_{xx}(\omega) &  \chi_{xy}(\omega) & 0\\  \chi_{yx}(\omega) & \chi_{yy}(\omega) & 0\\ 0 & 0 & 0\\ \end{array}\right), \nonumber \\
\chi_{xx}(\omega)&=&\frac{q_x^2\eta}{m\epsilon_0}\frac{A_y(\omega)}{A_x(\omega)A_y(\omega)-\sigma^2} ,\nonumber \\
\chi_{yy}(\omega)&=&\frac{q_y^2\eta}{m\epsilon_0}\frac{A_x(\omega)}{A_x(\omega)A_y(\omega)-\sigma^2} ,\nonumber\\
\chi_{yx}(\omega)=\chi_{xy}(\omega)&=&\frac{q_xq_y\eta}{m\epsilon_0}\frac{\sigma}{A_x(\omega)A_y(\omega)-\sigma^2} .
\end{eqnarray}
\begin{figure*}
\includegraphics[width=15cm]{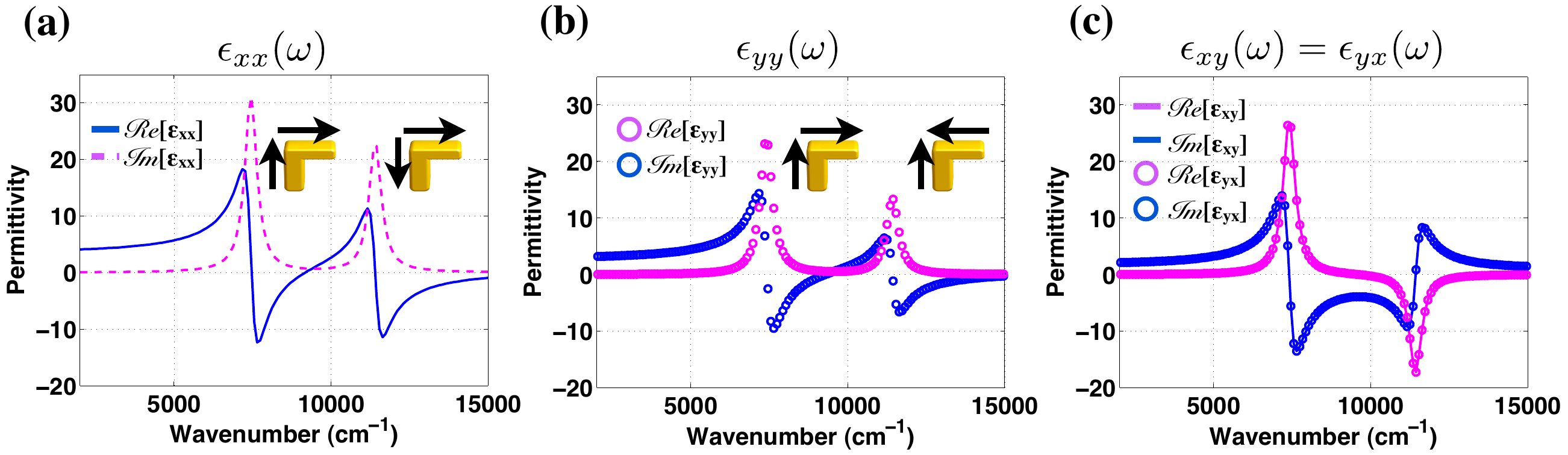}
\caption{(Color online) The diagonal effective permittivity tensor elements of the L -
metaatom: (a) $\epsilon(\omega)_{\mathrm{xx}}$ and (b)
$\epsilon(\omega)_{\mathrm{yy}}$. The arrows indicate the current flow,
given by direction of the carrier oscillation  for the particular eigenmode.
(c) The off-diagonal elements $\epsilon(\omega)_{ij}$ comprising a
Lorentz resonance around $\bar{\nu}=7,000~cm^{-1}$ and an anti-Lorentz
resonance at $\bar{\nu}=11,000~cm^{-1}$.} \label{pic4}
\end{figure*}
The most significant change compared to the SRR is the appearance of
off-diagonal elements in the susceptibility tensor $\hat{\chi}(\omega)$ which is, however, symmetric
leading also to
$\varepsilon_{ij}(\omega)=\varepsilon_{ji}(\omega)$. This symmetry
relation is important because it is required for time reversal, known
as the Onsager-Casimir principle \cite{Onsager1931,
Casimir1945,Tretyakov2002}. As expected the tensor of the effective permittivity has
the same form as that for planar optical active media
\cite{Zhukovsky2009,Bai2007} resulting in asymmetric transmission due
to elliptical dichroism. Note that the optical response would change
dramatically if both arms would be identical. In this case the
diagonal elements $\chi_{ii}(\omega)$ are identical too and the tensor
can be diagonalized by an rotation of $\pi/4$. So the polarization
eigenstates would be linear and the effective medium would be linearly
dichroitic. This is clear since the metaatom would have an additional
mirror symmetry with respect to the plane defined by the surface normal
and the line $x~=~y$, see Ref. \onlinecite{Wegener2009b}.

Another difference while comparing the L -  to the SRR metaatom is that
the splitting $\sigma$ between both resonances is reduced (by a factor
of $\sqrt{2}$), which follows from dropping one of the SRR arms.

In order to check whether our simple description is valid and to reveal the
relation between the SRR and the L - structure eigenmodes, we
performed again numerical FMM simulations for the L - metaatom and
compared the results to the analytical model in Fig. \ref{pic3}. Since we
deal now with more complex effective media where the full tensorial
nature of the permittivity tensor has to be taken into account the
standard transfer matrix algorithm cannot be applied. Hence, it is challenging
 to determine the scattering coefficients analytically
\cite{Borzdov1997} or even to invert them to retrieve effective
parameters directly. Therefore we used an adapted Fourier Modal
Method to determine the transmitted and reflected intensities
\cite{Lie2003}. In a first approximation we used the parameters which
we determined for the SRR. Based on these parameters we calculated
$\chi_{ij}(\omega)$ and the far-field intensities as shown in Fig.
\ref{pic3}(a,b) for the L - structure. The associated eigenmodes for
the carriers are shown in Fig. \ref{pic3}(c,d). The curves for both the
co-polarized Fig. \ref{pic3}(a,b) and the cross-polarized intensities Fig.
\ref{pic3}(e,f) are in good qualitative agreement for the rigorous FMM
results (dotted line) and the analytical model (dashed line) based on the
previously derived SRR oscillator parameters. Although there are
deviations between the actual resonance strength, the agreement, e.g.
for the resonance positions, is obvious. Note that the intensities, in
particular for the cross-polarized fields, can be solely predicted by the
coefficients obtained from the SRR. There, an almost perfect agreement
is observed.

In a second step we adapted the oscillator parameters in order to fit the
exact calculations [solid lines in Fig. \ref{pic3}(a,b,e,f)] providing an
almost perfect coincidence with the numerical values \footnote{We
mention that the fitting procedure can be manually performed since the
parameters are distinct with respect to their spectral effect, e.g.
resonance frequency ($\omega_0$), resonance splitting ($\sigma$),
resonance width ($\gamma$) and resonance strength
($q^2\eta/\epsilon_0m$).}. Note that the fitting is done only for the
co-polarized intensities from which the cross-polarized intensities follow.

A last step yields the effective permittivity tensor, that is inherently
accessible and already applied in order to fit the spectra in Fig. \ref{pic4}.
It can be seen that for the two polarization directions two eigenmodes
appear as Lorentz-shaped resonances for the effective permittivities.
They differ in strength, due to the different geometrical parameters of the
L -  arms, Fig. \ref{pic4}(a,b). The off-diagonal elements
$\epsilon_{ij}(\omega)$ are identical as discussed before, [Fig.
\ref{pic4}(c)]. Considering especially the second resonance for the
off-diagonal tensor elements near $\bar{\nu}=11,000~cm^{-1}$ , we
observe a Lorentzian anti-resonance that might suggest gain within the
system due to the negative imaginary part. However, the nature of this
resonance can be explained by the introduced
formalism as well. Since the permittivity is proportional to the
susceptibility Eq. (\ref{Eq4}) and hence also to the carrier displacements Eqs.
(\ref{Eq11}), a negative sign corresponds to a phase difference of $\pi$
between both oscillating carrier densities (antiparrallel oscillations), while for the positive Lorentz resonance
at around $\bar{\nu}=7,000~cm^{-1}$, both are oscillating in-phase
[sketched by the arrows in Fig. \ref{pic4}(a)]. Thus, the negative sign in the imaginary part of the permittivity can by fully explained by means of the mutual interplay of the coupled oscillators.

\section{The S -  metaatom}
\begin{figure*}
\includegraphics[width=15cm]{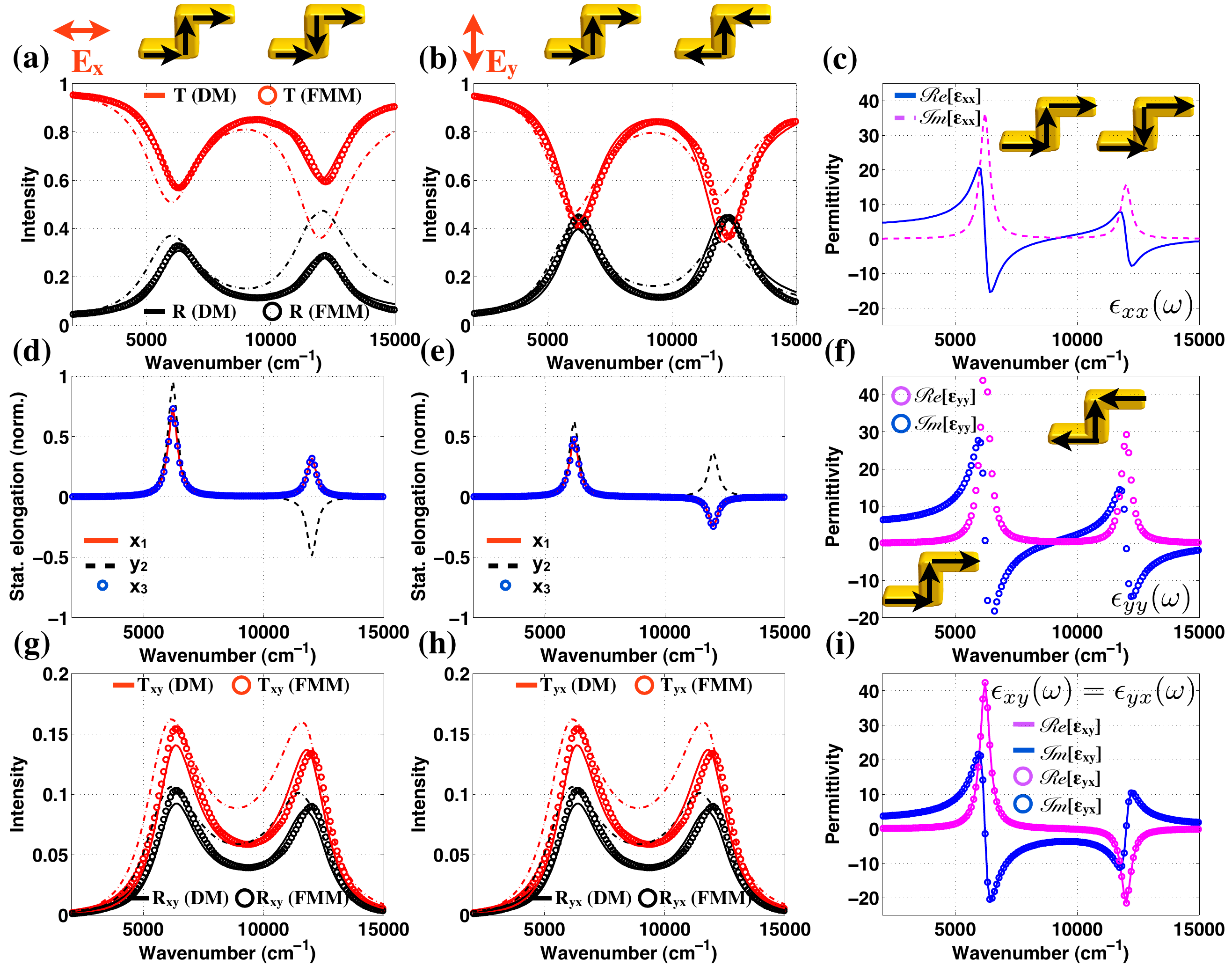}
\caption{(Color online) The far field spectra of the S - structure (a) $x$-polarization and
(b) $y$-polarization obtained by numerical simulations (circles),
predictions based on the SRR structure parameters (dashed-dotted lines),
and adapting the parameters to fit the numerical values (solid lines). The
carrier eigenmodes, i.e., an in line current over the entire structure and
antiparallel currents [normalized imaginary part of $x_1(\omega),
y_2(\omega), x_3(\omega)$] with respect to the center part of the S -
structure are observed for the two polarization directions (d) $x$ and (e)
$y$. The comparison between the cross-polarization contributions
($T_{ij},~R_{ij}$) for the numerical simulations (circles), the predicted
lines from the SRR parameters (dashed-dotted lines) and the fitted
parameter spectra (solid lines) for the found parameters deduced for
fitting the co-polarized response ($T_{ii},~R_{ii}$). The effective
permittivity tensor diagonal (c), (f) and the off-diagonal components (i).}
\label{pic5}
\end{figure*}

In order to verify the model we are going to investigate another modification of
the SRR, namely the S - structure \cite{Chen2005_1,Chen2005_2} [Fig.
\ref{pic1}(a)].  To observe optical activity with the same number of
coupled entities as for the SRR its mirror symmetry has to be broken.
Therefore one of the SRR arms (e.g. $x_3$) is turned with respect to the
SRR base.

This {\it opening} of the SRR structure is expected to enable the
observation of two modes for $x$-polarization, since the oscillator in the
base ($y_2$) can now oscillate in-phase or out-of-phase with the two
remaining $x$-oriented oscillators of the S - structure that are excited by
the $x$-polarized electric field. For $y$-polarization the situation is
similar, but now the two oscillators in the horizontal arms ($x_1,~x_3$)
are allowed to oscillated in-phase or out-of-phase to the excited oscillator
in the base ($y_2$). Mathematically this modification can be considered by
setting $\sigma_{23}=-\sigma_{21}\equiv \sigma$ (see Eq. \ref{Eq1}),
while all other parameters appear similar to the ones applied for the SRR.


With these initial assumptions, which reflect all modifications to the
geometry, the calculations can be repeated in analogy to those for the
SRR and the L -  structure. Thus, we obtain the elongations for
$x$-polarized excitation
\begin{eqnarray}
\label{Eq12}
x_1^x&=&x_3^x=-\frac{q_x}{m}\frac{A_y(\omega)}{A_x(\omega)A_y(\omega)-2\sigma^2}E_x(z,\omega) , \nonumber \\
y_2^x&=&-\frac{q_x}{m}\frac{2\sigma}{A_x(\omega)A_y(\omega)-2\sigma^2}E_x(z,\omega),
\end{eqnarray}
and for $y$-polarized excitation
\begin{eqnarray}
\label{Eq13}
x_1^y&=&x_3^y=-\frac{q_y}{m}\frac{\sigma}{A_x(\omega)A_y(\omega)-2\sigma^2}E_y(z,\omega) , \nonumber \\
y_2^y&=&-\frac{q_y}{m}\frac{A_x(\omega)}{A_x(\omega)A_y(\omega)-2\sigma^2}E_y(z,\omega),
\end{eqnarray}
as well as the respective effective susceptibility tensor
\begin{eqnarray}
\label{Eq14}
\chi(\omega)&=&\left(\begin{array}{ccc} \chi_{xx}(\omega) &  \chi_{xy}(\omega) & 0\\  \chi_{yx}(\omega) & \chi_{yy}(\omega) & 0\\ 0 & 0 & 0\\ \end{array}\right), \nonumber \\
\chi_{xx}(\omega)&=&\frac{q_x^2\eta}{m\epsilon_0}\frac{2A_y(\omega)}{A_x(\omega)A_y(\omega)-2\sigma^2} ,\nonumber \\
\chi_{yy}(\omega)&=&\frac{q_y^2\eta}{m\epsilon_0}\frac{A_x(\omega)}{A_x(\omega)A_y(\omega)-2\sigma^2} ,\nonumber\\
\chi_{yx}(\omega)=\chi_{xy}(\omega)&=&\frac{q_xq_y\eta}{m\epsilon_0}\frac{2\sigma}{A_x(\omega)A_y(\omega)-2\sigma^2}.
\end{eqnarray}
In Eqs. (\ref{Eq14}) we have used the polarization Eq. (\ref{Eq2}) which is
found to coincide with that of the SRR structure, whereas the elongations
are different. With respect to the splitting of the resonances, we expect
the same resonance positions as for the SRR for $y$-polarization, since
$2\sigma$ appears in the denominator of the SRR oscillation amplitudes
[Eqs. (\ref{Eq7}) for $y$-polarization] as well as in all elongations in Eqs. (\ref{Eq12}) and Eqs. (\ref{Eq13}).

Performing the respective numerical and analytical calculations as before
for the L - structure, we can predict the spectral response as well as the
effective material properties based on the plasmonic eigenmodes. The
results are shown in Fig. \ref{pic5}.

Considering these eigenmodes in Fig. \ref{pic5}(d,e), we may distinguish
two situations. At first an eigenmode which is represented by three
dipoles being in phase along the entire structure. The second eigenmode
is characterized by two dipole being in-phase in the arms and
out-of-phase in the base. Both eigenmodes are excited for $x$- and
$y$-polarization, respectively, and will lead to spectral resonances
appearing as dips and peaks in transmission or reflection, respectively
[Fig. \ref{pic5}(a,b)]. The spectral positions of the resonances are in
agreement with the expectations from the SRR structure resonances.

As expected, for both the S -  and the L - metaatom the in-phase
eigenmodes appear at smaller wavenumbers (larger wavelengths)
compared to the out-of-phase ones. This is completely in agreement with
arguments from plasmon hybridization theory \cite{Nordlander2003}.

Again the use of the oscillator parameters as optimized for the SRR
structure [dash-dotted lines in Fig. \ref{pic5}(a,b,g,h)] reveals the
relationship between both structures, since the numerically (circles)
calculated spectra agree with respect to the overall shape and the
resonance positions Fig. \ref{pic5}(d,e) very well with the analytical
predictions (solid and dashed lines). A subsequent fitting again improves the results towards
almost excellent agreement, which can also be observed for the cross-
polarization observables Fig. \ref{pic5}(g,h). Considering the tensor
components of the effective permittivity, we observe a difference
between the diagonal entries due to the geometrical differences in the S -
structure center and arms, while the anti-resonance for the out-of-phase
eigenmode is observed in the off-diagonal elements with the same origin
as discussed for the L-structure.

Finally we provide all parameters required for the analytical calculations
presented here for the SRR as well as the two presented modifications in
Tab. \ref{tab1}.
\begin{table}
\caption{The fitted oscillator parameters applied to reproduce the optical
far-field intensities for the SRR as well as the adapted parameters for the
L-  and S- structure are presented. For consistence with the
corresponding figures, all spectral units are given in wavenumber
units ($cm^{-1}$).}
\begin{center}
\begin{tabular}{lccc}
\hline\hline
                            Fitting parameter   &   SRR         &   {\it L}         &    {\it S}    \\ \hline
$\omega_{0x}, \omega_{0y} ~ [cm^{-1}]$          &   $9770$,$~9050$  &   $9770$,$~9500$  &   $9770$,$~9350$  \\
$\gamma_x, \gamma_y ~ [cm^{-1}]$                &   $520$,$~420$        &   $520$,$~420$        &   $520$,$~420$        \\
$\sqrt{\sigma} ~ [cm^{-1}]$                 &   $6100$          &   $6100$          &   $6100$          \\
$q_x^2\eta / \epsilon _0m~ [AsV/m^2Kg]$&    $0.65\cdot 10^{39}$ &   $0.83\cdot 10^{39}$ &   $0.35\cdot 10^{39}$  \\
$q_y^2\eta / \epsilon _0m ~ [AsV/m^2Kg]$&   $1.1\cdot 10^{39}$  &   $0.55\cdot 10^{39}$ &   $1.10\cdot 10^{39}$ \\
\hline\hline
\end{tabular}
\end{center}
\label{tab1}
\end{table}

\section{Summary}
In summary, we have presented an analytical model which permits the calculation
of effective material parameters for planar metamaterials consisting of variable metaatoms formed by 
a few straight wire sections of potentially different shape.
The model takes advantage of resonant electric dipole oscillations in the wires and their mutual
coupling. The pertinent form of the metaatom determines the actual coupling features.
Thus this model represents a kind of building block approach for quite different metaatoms.
 Once 
the constants describing the respective dipole oscillations
for one particular arrangement  have been determined, where here the SRR has been used, the properties
of another, modified metaatom can be easily predicted. 
Since in particular the effect of asymmetric transmission for
circular polarized light attracted a lot of research interest recently, we
focused here on planar metaatoms that are optically active. Within our
model all properties of the effective material tensors for such kind of
media are correctly predicted  and the corresponding scattering
characteristics are in very good agreement with the rigorous numerical
results.

\section{Acknowledgements}
Financial support by the Federal Ministry of Education and Research (ZIK,
MetaMat) as well as from the State of Thuringia within the ProExcellence
program is acknowledged.

\bibliography{PRB_FL}
\end{document}